\documentclass[vecphys]{svmult}

\usepackage{makeidx}         
\usepackage{graphicx}        
\usepackage{multicol}        
\usepackage[bottom]{footmisc}

\makeindex             

\begin{document}

\title*{The quest for obscured AGN at cosmological distances: Infrared 
Power-Law Galaxies}
\titlerunning{Infrared Power-Law Galaxies} 
\authorrunning{Alonso-Herrero et al.}
\author{Almudena Alonso-Herrero\inst{1,2},
Jennifer L. Donley\inst{2},
George H. Rieke\inst{2}, Jane R. Rigby\inst{3,2} \and Pablo
G. P\'erez-Gonz\'alez\inst{4,2}}
\institute{DAMIR, Instituto de Estructura de la Materia, CSIC, 
28006 Madrid, Spain
\texttt{ }
\and Steward Observatory, University of Arizona, Tucson, AZ 85721, USA
\texttt{}
\and Carnegie Observatories, Pasadena, CA 91101, USA
\texttt{}
\and Departamento de
Astrof\'{\i}sica y CC de la Atm\'osfera, UCM, 28040 Madrid, Spain\texttt{}
}
%
%
\maketitle
\index{Author1}
\index{Author2}


\begin{abstract}
We summarize multiwavelength properties of a sample of galaxies
in the {\it Chandra} Deep
Field North (CDF-N) and South (CDF-S) whose Spectral Energy Distributions
(SEDs) exhibit the characteristic power-law behavior expected for AGN in the
{\it Spitzer}/IRAC $3.6-8\,\mu$m bands. AGN selected this way tend to comprise
the majority of high X-ray luminosity AGN, whereas AGN selected via other IRAC color-color criteria might contain more star-formation dominated
galaxies. Approximately half of these IR power-law
galaxies in the CDF-S are detected in deep (1\,Ms) {\it Chandra} X-ray imaging,
although in the CDF-N (2\,Ms) about 77\% are detected at the
$3\,\sigma$  level.  The SEDs and X-ray upper limits
of the sources not detected in X-rays are consistent with
those of obscured AGN, and are significantly different
from those of massive star-forming galaxies. About
40\% of IR power-law galaxies detected in X-rays have SEDs
		    resembling that of an optical QSO and morphologies dominated by  
bright point source emission.  The remaining 60\% have SEDs whose UV and
		    optical continuum are much steeper (obscured) and more extended
		    morphologies than those detected in X-rays.
 Most of the IR power-law galaxies not
		    detected in X-rays  have IR ($8-1000\,\mu$m)
		    above $10^{12}\,{\rm L}_\odot$, and X-ray (upper limits)
		    to mid-IR ratios 
		    similar to those of local warm (ie, hosting an AGN)
		    ULIRGs. The SED shapes of power-law galaxies 
		    are consistent with the obscured fraction (4:1) as
		    derived from the X-ray column densities, if we assume
		    that all the sources not detected in X-rays are heavily
		    absorbed. IR power-law galaxies may account for 
between 20\% and 50\% of the predicted number density of mid-IR detected
obscured AGN. The remaining obscured 
AGN probably have rest-frame SEDs dominated by stellar emission.
\end{abstract}

\section{Introduction}
\label{sec:intro}
Active Galactic Nuclei (AGN) are 
sources of luminous X-ray emission, and at cosmological distances AGN
are routinely selected from deep X-ray ($<10\,$keV) exposures
(\cite{brandt}). 
Highly obscured ($N_{\rm H} > 10^{23}-10^{24}\,{\rm cm}^{-2}$) AGN
are thought to be a major contributor to the hard X-ray background
(\cite{mushotzky,comastri,worsley04,worsley05}). However, the majority of them 
might not be detected in
these X-ray surveys because a large fraction of their soft X-ray, UV, and
optical emission is absorbed, and presumably reradiated in the infrared (IR).

 Numerous attempts have been made to detect this population of heavily
 obscured AGN, many of which have focused on the MIR emission where the
 obscured radiation is expected to be reemitted (e.g.,
 \cite{aah06,lacy04,stern05,polletta06,donley06}), or on 
 combinations of MIR and multiwavelength data (e.g.,
 \cite{donley05,martinez05}). In the mid-IR, AGN can often be distinguished
 by their characteristic power-law emission (e.g.,\cite{neugebauer,elvis}). 
This
 emission is not necessarily due to a single source, but can arise from the
 combination of non-thermal nuclear emission and thermal emission from various
 nuclear dust components (\cite{riekelebofsky}). We summarize here 
the properties of galaxies showing
 the characteristic power-law behavior expected for AGN in the {\it Spitzer}
 $3.6-8\,\mu$m bands detected in the {\it Chandra} Deep Field North and South 
(CDF-N and CDF-N) studied in \cite{aah06} and \cite{donley06}, 
respectively. We also discuss results from other IR-based methods to detect
 high-$z$ obscured AGN. We use $H_0 = 71\,{\rm km \,s}^{-1}\,{\rm Mpc}^{-1}$, 
$\Omega_{\rm M} = 0.3$, and $\Omega_\Lambda = 0.7$.

\section{The Sample of IR power-law galaxies}

\subsection{Selection}

We chose two cosmological fields with
deep X-ray coverage (CDF-N: 2\,Ms and CDF-S: 1\,Ms, see \cite{alexander} for
details) to look for obscured AGN. 
We selected as power-law galaxies sources that were detected 
in each of the four IRAC 
($3.6, \, 4.5, \, 5.8,$ and $8\,\mu$m) bands and whose
IRAC spectra could be fitted as $f_\nu \propto \nu^\alpha$, where $\alpha$ 
is the 
 spectral index. We used a minimum $\chi^2$ criterion to select 
galaxies whose IRAC SEDs followed a power law with 
spectral index $\alpha < -0.5$. The choice for the spectral index was based on
the empirical spectral energy distributions (SEDs) 
of bright QSO selected in the optical, X-rays and near-IR 
(e.g., \cite{neugebauer,elvis,ivezic,kura}) and
Seyfert galaxies (e.g., \cite{ward,edelson}).
There are two slight differences in the two catalogs of IR power-law
galaxies. In the CDF-S \cite{aah06} 
started the selection of the power-law candidates 
from the $24\,\mu$m catalog of
\cite{papovich}, without any further requirements on the flux limits of the
IRAC catalogs. \cite{donley06} in the CDF-N instead imposed a strict S/N$=6$
flux density 
cut in each of  the IRAC bands but did not require a $24\,\mu$m detection,
although virtually all of the power-law galaxies in the CDF-N were also detected 
at $24\,\mu$m down to $\sim 80\,\mu$Jy (equivalent to the 80\% completeness
limit of the CDF-S, see \cite{papovich}). 

To minimize the chances of selecting non-active
galaxies we constructed optical-MIR SEDs (see
\S4), and compared them with theoretical
and observational templates of star-forming galaxies. 
We rejected any source selected via
the power-law criteria whose SED resembled a
star-forming galaxy. The final samples included 92 and 
62 galaxies in the CDF-S and CDF-N, respectively 
(see \cite{aah06} and \cite{donley06} for details).

A small ($\sim 25-30\%$, depending on the field) 
fraction of the power-law galaxies, typically
the optically bright X-ray sources (see next section), 
have spectroscopic redshifts (e.g.,
\cite{szokoly} in CDF-S). We supplemented the available spectroscopic
redshifts with photometric ones estimated with an improved version of the
method described by \cite{perezgonzalez}. We find that the IR power-law
galaxies tend to lie at significantly higher redshifts ($z>1$)  than the X-ray
sources  (median $z\sim 0.7$, see \cite{brandt} for a review) in both
fields. 

\begin{figure}[!t]
\parbox[t]{\textwidth}{%
\vspace{0pt}
\includegraphics[width=5.9cm]{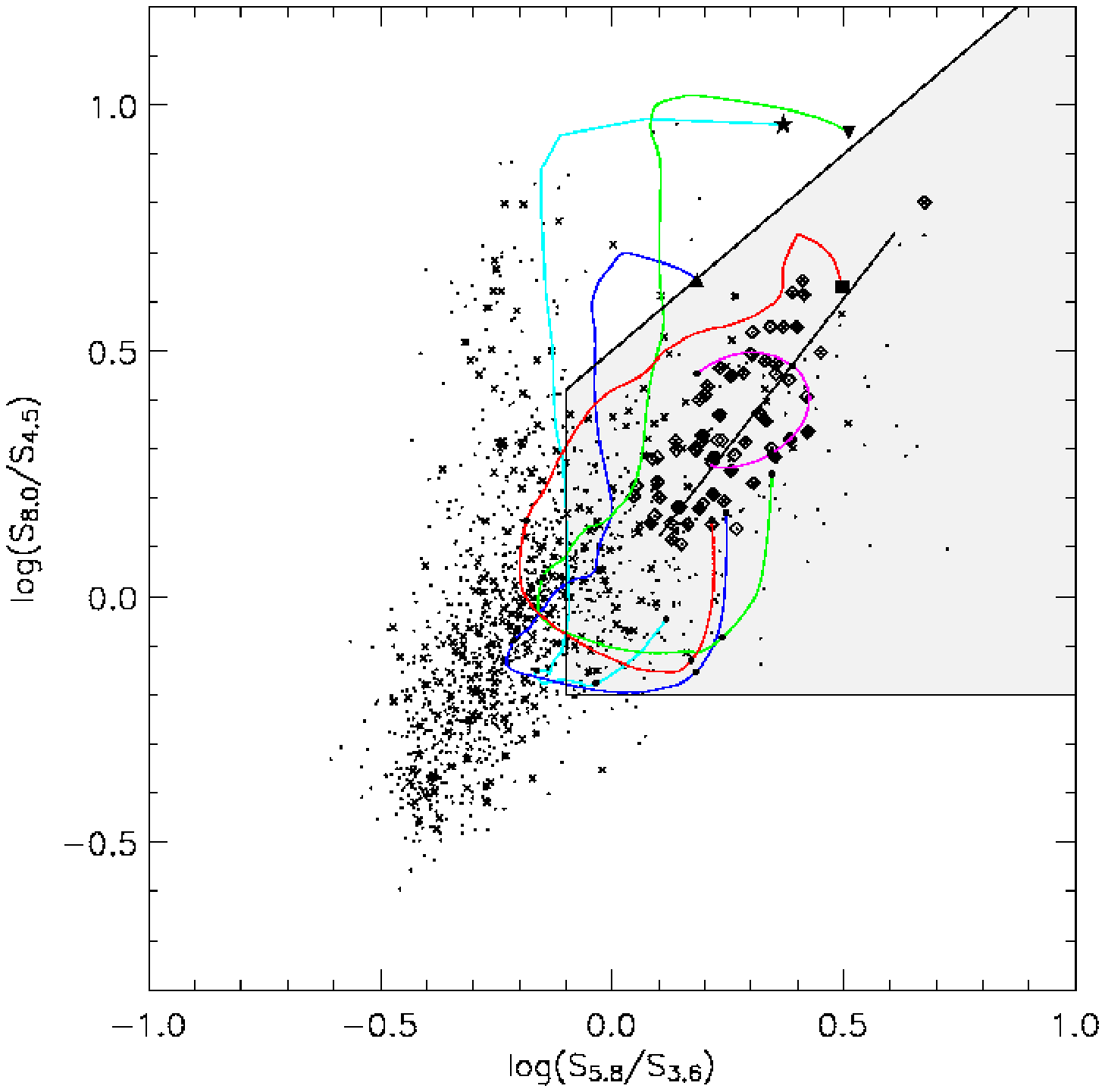}
\hfill
\includegraphics[width=5.9cm]{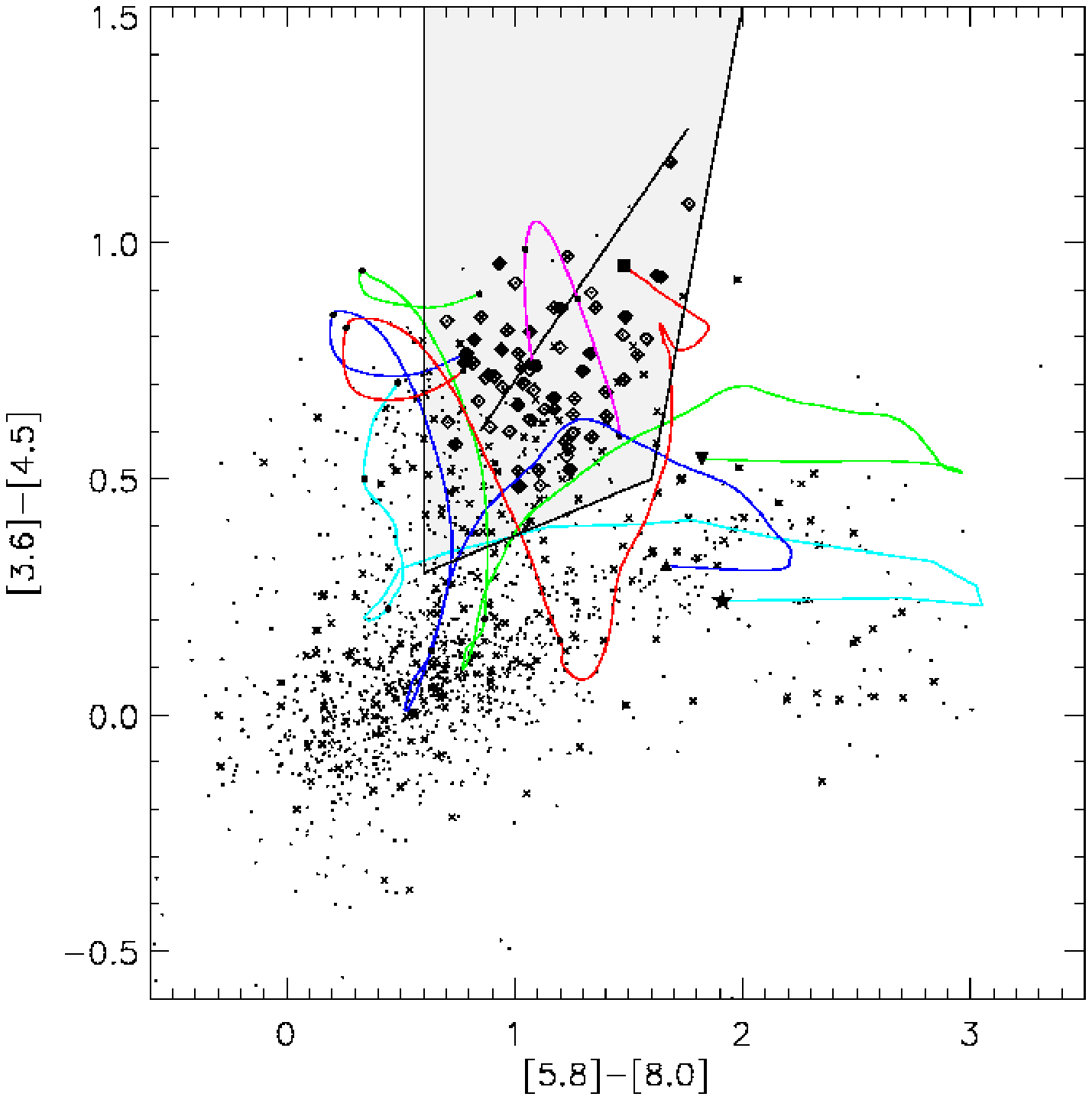}}
\vspace{-1.5cm}

%
%
\caption{Location of the CDF-N IR power-law galaxies (diamonds) on the Lacy et
al. (left) and Stern et al. (right) IRAC color-color diagrams together with
X-ray sources (crosses from \cite{alexander}) and IRAC galaxies (dots, 
non power-law galaxies) detected in this field. The shaded
regions indicate the AGN loci, and the straight lines within them the
power-law criterion. We also show the $z$-evolution ($z=0-2.5$ where 
$z=0$ is indicated by the large symbol at the edge of each template 
line) of different 
templates. In green (upside down triangle) a starburst ULIRG, in blue (triangle)
Arp~220, in red (square) an AGN ULIRG,
in pink (filled dot) the average of the radio-quiet QSOs of \cite{elvis}, and
in cyan (star symbol) a star-forming galaxy.}
\label{fig:1}       
\end{figure}

\subsection{Comparison between IRAC power-law and color-color criteria}

\cite{lacy04} and \cite{stern05} 
defined AGN selection criteria based on {\it Spitzer}/IRAC
color-color diagrams. The Lacy et al criterion is based on SDSS QSOs, and
therefore excludes 
AGN in which the host galaxy dominates the MIR (see e.g.,
\cite{aah04,rigby06,franceschini,polletta06}), 
as well as AGN obscured in the
mid-IR. Our power-law galaxies fall along a straight 
line well within the Lacy et al diagram, although they do not cover completely 
the available color space (see Fig.~1). The
Stern et al criteria are based on the observed properties of 
spectroscopically classified AGN, and provide a closer
match to our power-law technique. While the color-color selected samples
comprise a higher fraction of the low X-ray luminosity AGN than does the
power-law selected sample, the color criteria select more sources not detected
in X-rays, due at least in part to a higher degree of contamination from
ULIRGs dominated by star formation (Fig.~1 and \cite{barmby}). 

\begin{figure}[!t]
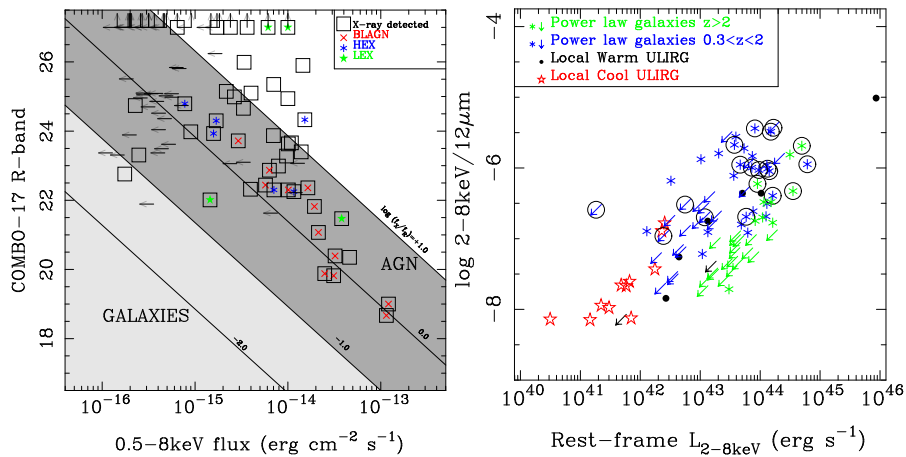

\parbox[t]{\textwidth}{%
\vspace{0pt}
\includegraphics[width=5.8cm]{plen_talk_alonsoherrero_f2a.ps}
\hfill
\includegraphics[width=6.cm]{plen_talk_alonsoherrero_f2b.ps}}

%
%
\caption{{\it Left panel:} COMBO-17 (\cite{wolf}) 
$R$-band versus the {\it Chandra} full band
  flux of IR power-law galaxies in the CDF-S (similar results are found for
  the CDF-N). When available we show the
  optical   spectroscopic classifications from \cite{szokoly} (see \S5 for
  details). The galaxies in our sample not detected by COMBO-17 are shown as
  upper limits at $R=27\,$mag. {\it Right panel:} Rest-frame $2-8\,$keV
  luminosity vs hard X-ray to $12\,\mu$m for IR power-law galaxies (asterisks
  and X-ray upper limits; the circles denote those classified as BLAGN) 
in the CDF-S
  compared with local cool (no AGN) and warm (hosting an AGN) ULIRGs.  }
\label{fig:2}       
\end{figure}

Recently \cite{lacy06} obtained follow-up optical spectroscopy of 
objects selected according to \cite{lacy04} in the {\it Spitzer}
First Look Survey and SWIRE XMM-LSS fields. Their sample is flux-limited at
$24\,\mu$m, although their objects  are 
much brighter ($f_{24\mu{\rm m}} = 4-20\,$mJy, median of 5\,mJy, and median
$R$-band magnitudes $R\sim 18\,$mag) than our power-law galaxies
($f_{24\mu{\rm m}} \sim 
0.08-3\,$mJy, and $R\sim 23\,$mag for those detected by COMBO-17 see next
section), and on average their AGN are closer   
$z_{\rm sp} \sim 0.6$ compared with $z\sim 1.5$ for our power-law galaxies. 
The location of this sample on
the \cite{lacy04} IRAC color-color diagram (figure~7 in \cite{lacy06}) is
almost identical to the positions of the power-law galaxies shown in Fig~1
(right panel). Their selection technique has proven to be very effective at 
selecting AGN as their follow-up spectroscopy shows that approximately 90\%
have  AGN signatures with one-third of them showing broad-line regions, thus an
obscured-to-unobscured ratio of 2:1 (see \S5 for the power-law galaxies). All
these properties seem to indicate that  these color-color selected galaxies 
represent the brightest end of the power-law galaxies.

\section{X-ray, Infrared, and Optical Properties} 

In both CDF-N and CDF-S we found that approximately 50\% of the 
IR power-law galaxies were detected in at least one {\it Chandra} band 
using the X-ray catalogs of \cite{alexander}. In the CDF-S 
we stacked the X-ray data of a few individually 
undetected galaxies (at off-axis angles of $\theta <7.5'$) 
and found a significant detection in the hard-band
($3.1\,\sigma$) and a tentative detection in the soft band ($2\,\sigma$). For  
$z=2$ these would correspond to observed soft and hard luminosities of $<7
\times 10^{41}\,{\rm erg \, s}^{-1}$ 
and $4\times 10^{42}\,{\rm erg \, s}^{-1}$, respectively. This is consistent
with obscured AGN. Since the X-ray exposure of the CDF-N is twice that of
CDF-S we  searched for faint X-ray emission at the positions of the power-law
galaxies not in the \cite{alexander} catalog. We found that the X-ray
detection rate 
increases to 77\% at the $3\,\sigma$ level. 
The power-law galaxies make up a
significant fraction of the high X-ray luminosity sample, as 
our selection criteria
require the AGN to be energetically
dominant. The lower luminosity X-ray sources not
identified as power-law galaxies tend to be
dominated by the $1.6\,\mu$m stellar bump in the optical to near-IR bands (see also
\cite{aah04,franceschini,polletta06,rigby06}). 

\begin{figure}[!t]
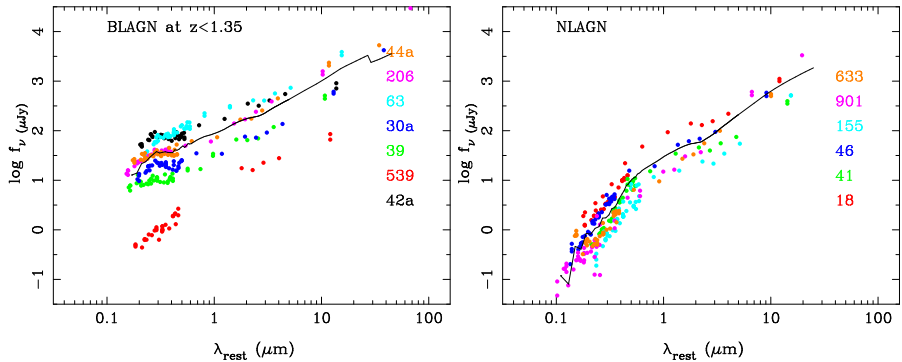

\parbox[t]{\textwidth}{%
\vspace{0pt}
\includegraphics[width=5.9cm]{plen_talk_alonsoherrero_f3a.ps}
\hfill
\includegraphics[width=5.9cm]{plen_talk_alonsoherrero_f3b.ps}}

%
%
\caption{Rest-frame SEDs (filled circles) of CDF-S IR power-law galaxies
  detected in 
  X-rays with spectroscopic redshifts and classifications from \cite{szokoly} 
  (IDs given on the right-hand side of each panel). For each SED class  (only
  the BLAGN and NLAGN classes are shown, see \cite{aah06} for more details) we have
  constructed an average template, shown as the solid line in each panel.}
\label{fig:3}       
\end{figure}

A large fraction of IR luminous high-$z$ galaxies 
have been  found to host AGN (e.g., SCUBA galaxies \cite{alexander05}), and
IR luminous galaxies at $z \sim 1-2$ (e.g., \cite{yan}).
We measured the total IR ($8-1000\,\mu$m) luminosity of the CDF-S 
power-law galaxies from the rest-frame $12\,\mu$m  
luminosity. Although our procedure to compute
IR luminosities is  similar to that of \cite{perezgonzalez},
we took special care to use $12\,\mu$m to IR luminosity ratios specific to the
class of galaxies in study. In particular, galaxies whose SEDs resemble those
of optical QSOs (see \S4 and Fig.~3) 
show $12\,\mu$m to IR luminosity ratios significantly 
lower than the typical
values of cool ULIRGs and some warm ULIRGs (e.g., Mrk~231). 
All the IR power-law galaxies are highly luminous.  About 30\% are in the
hyperluminous class ($L_{\rm IR} > 10^{13}\,{\rm L}_\odot$), 41\% are ULIRGs 
($L_{\rm IR} = 10^{12}-10^{13}\,{\rm L}_\odot$), and all but one of the rest
are LIRGs ($L_{\rm IR} = 10^{11}-10^{12}\,{\rm L}_\odot$).  
At the lower 
IR luminosity end ($L_{\rm IR} < 10^{12}\,{\rm L}_\odot$) a large fraction
are detected in X-rays and tend to have SEDs similar to those of optical QSOs
(see next section). 

\begin{figure}[!t]
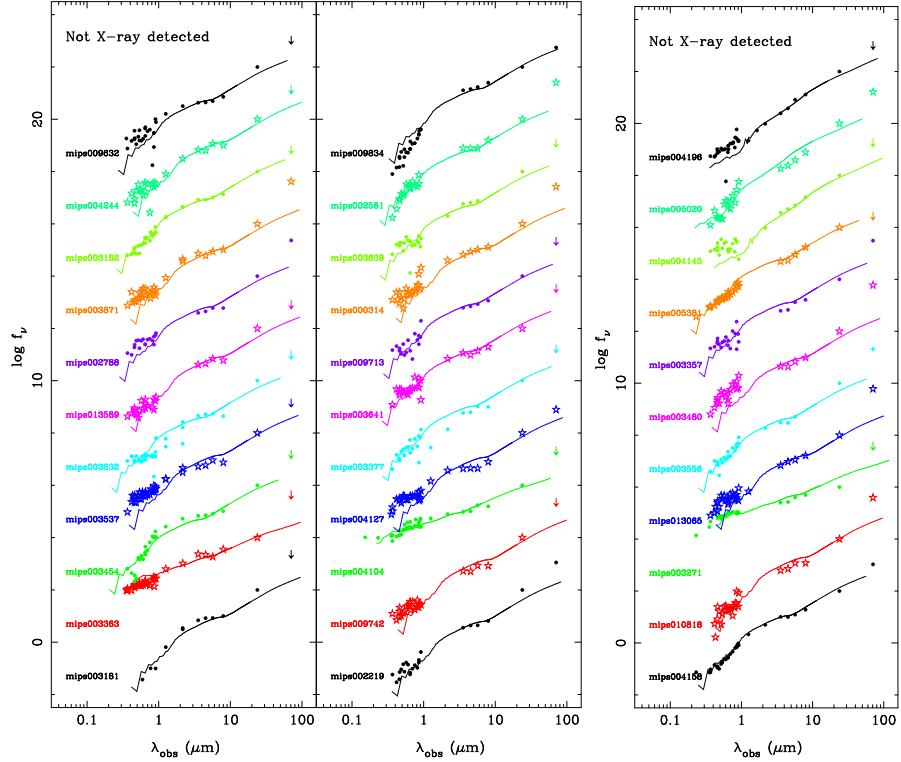

\parbox[t]{\textwidth}{%
\vspace{0pt}
\includegraphics[width=7.6cm]{plen_talk_alonsoherrero_f4a.ps}
\hfill
\includegraphics[width=4.05cm]{plen_talk_alonsoherrero_f4b.ps}}

%
%
\caption{Examples of the observed SEDs (filled circles and star symbols) of CDF-S IR power-law
  galaxies not detected in
  X-rays. Each galaxy is shown with the closest average template constructed
  using the X-ray detected ones (see
  \cite{aah06} for details).}
\label{fig:4}       
\end{figure}

About one-quarter of the CDF-S IR power-law galaxies are optically faint
i.e., were not detected by COMBO-17 (\cite{wolf}) 
down to a limit of $R\sim 26.5\,$mag. Moreover, the fraction of power-law
galaxies not detected in X-rays increases
toward fainter $R$-band magnitudes (see Fig.~2), an indication of their
obscured nature as 
also revealed by their SEDs (see also \S5). 
A number of works (e.g., \cite{bauer} and references therein) have
demonstrated that X-ray to optical flux ratios can be useful for
distinguishing between AGN and star-forming galaxies for sources detected in
deep X-ray exposures. Fig.~2 (left panel) 
shows that the majority of the galaxies (or
their upper limits) in the CDF-S  are consistent with being AGN or transition
objects based 
on the X-ray vs. $R$-band diagram (similar results are found for the CDF-N
galaxies).  The location
of the IR power-law galaxies on this diagram (see \cite{bauer}) 
indicates X-ray luminosities (or
upper limits) above $10^{41}\,{\rm erg \, s}^{-1}$, as also shown by the right
panel of Fig.~2. The
rest-frame hard X-ray/$12\,\mu$m ratios of the IR power-law
galaxies are similar to those of local warm
ULIRGs (i.e., those containing an AGN) and
QSO.

\begin{figure}
\includegraphics[width=11.8cm]{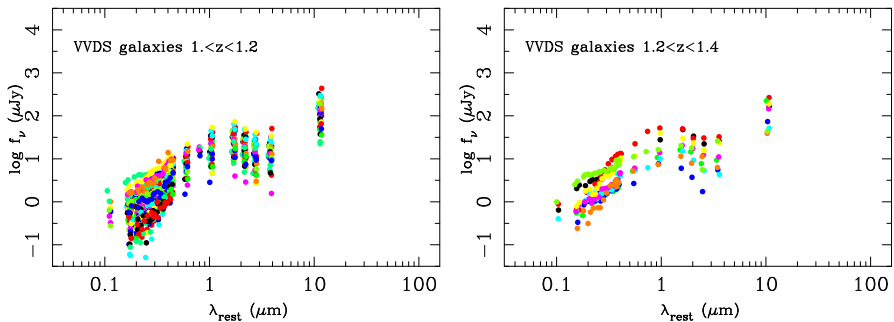}
\caption{Examples of rest-frame SEDs of CDF-S VVDS galaxies at $z>1$ included in
  the sample of predominantly star-forming galaxies selected at $24\,\mu$m by
  \cite{perezgonzalez}. }
\label{fig:5}       
\end{figure}

%
%
%

\section{SEDs and Morphologies}

Using available multiwavelength datasets (see \cite{aah06,donley06} for references)
of the two cosmological fields we constructed SEDs for our sample. The 
spectroscopic classifications (broad vs. narrow lines\footnote{The Szokoly et
  al. spectral  classifications of CDF-S X-ray sources with clear AGN signatures
  were: BLAGN (broad-line
AGN) and HEX  (high excitation lines). Approximately 50\% of their X-ray sources did not
have a clear AGN signature in their optical spectra: LEX (low excitation lines, also termed
optically-dull AGN and X-BONGS) and ABS (absorption lines).}) tend to agree with two
distinct types of SEDs.  
About 40\% of the CDF-S IR power-law galaxies detected in X-rays are
classified as BLAGN  and have SEDs
(Fig.~3) similar to the average radio-quiet QSO SED of \cite{elvis}, that is,
with an optical--to--mid-IR continuum almost flat in $\nu f_\nu$ with a UV
bump.   The remaining X-ray sources with narrow lines (NLAGN) 
have SEDs similar to the BLAGN but their UV and
optical continua are much steeper (obscured), and some of them resemble local
warm ULIRGs. The majority of
the power-law galaxies not detected in X-rays (Fig.~4) have steep SEDs
similar to the NLAGN or ULIRG class as they tend to be optically
fainter (see \S3) and possibly more obscured (see \S5) than the X-ray sources.  
\begin{figure}
\centering
\includegraphics[width=13cm]{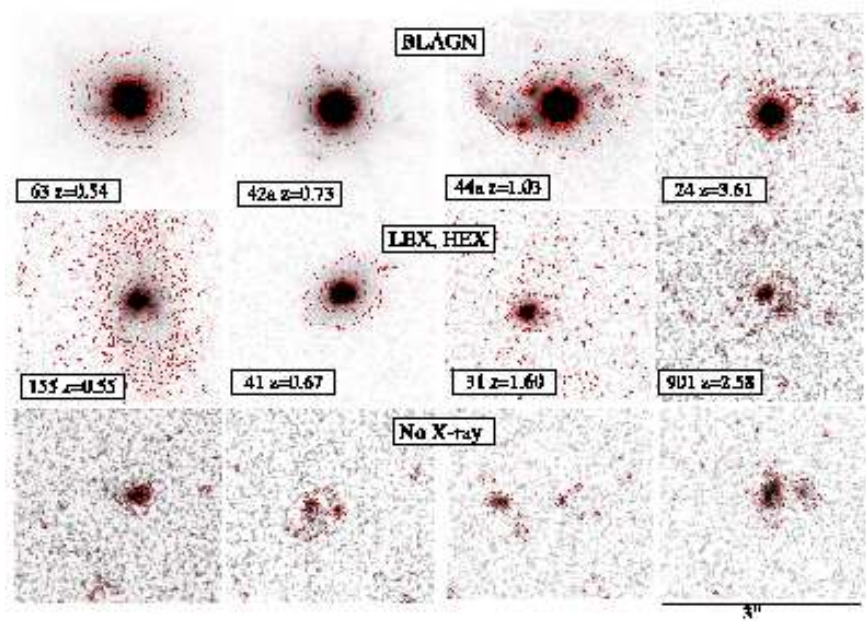}
%
%
\caption{Postage stamps of optically-bright power-law
galaxies in the CDF-S. The data are from the public release of
the GOODS {\it HST}/ACS F606W images. Examples
of X-ray detected power-law galaxies are shown in the top and middle panels
with the IDs, and spectroscopic 
redshifts and classifications from \cite{szokoly}. Power-law galaxies not
detected in X-rays are shown in the bottom panel. }
\label{fig:6}       
\end{figure}

In contrast with our power-law galaxies, massive galaxies from the
VIRMOS VLT Deep Survey (VVDS, \cite{lefevre})
at $1<z<2$ with $24\,\mu$m detections from the sample of
\cite{perezgonzalez} are predominantly
star-forming  galaxies with a prominent stellar bump at
$1.6\,\mu$m due to an evolved (red giants and supergiants) stellar population.
Moreover, the SEDs of power-law galaxies are also significantly different from
those of the majority of optically-dull AGN in the CDF-S 
which show SEDs dominated by stellar light originating in  
the host galaxy (see \cite{rigby06}).

\begin{figure}
\includegraphics[angle=-90,width=11.8cm]{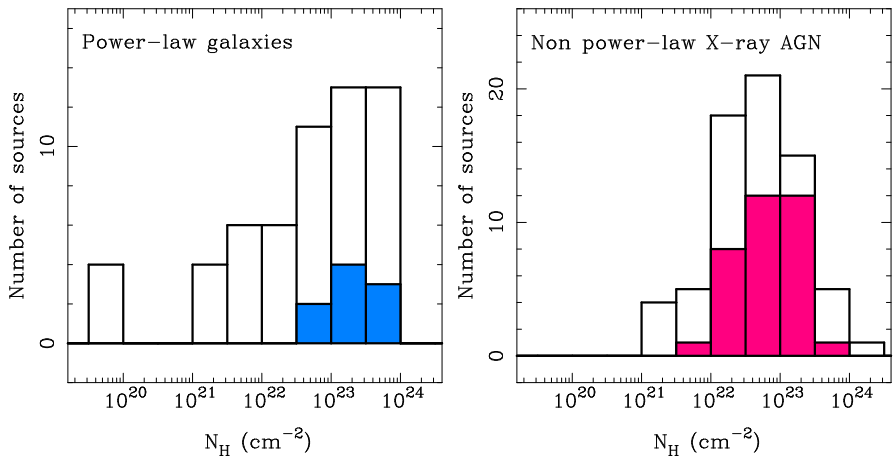}
\caption{{\it Left panel}. 
Distribution of X-ray column densities for the CDF-N power-law
  galaxies that are cataloged X-ray sources (empty) and 
those only weakly
  detected in X-rays (shaded) from \cite{donley06}. We also include 
those CDF-S power-law galaxies selected by \cite{aah06} for
  which \cite{rigby06} measured a column density.   
{\it Right Panel}. $N_{\rm H}$ distribution for all the 
  CDF-S  X-ray optically active AGN (empty) with spectroscopic classifications 
  \cite{szokoly,rigby06}, excluding all 
the power-law galaxies shown in the left
  panel. The shaded histogram shows the $N_{\rm H}$ of 
the optically-dull AGN (only those classified as LEX).} 
\label{fig:7}       
\end{figure}

Fig.~4 shows the Great Observatories Origins Deep 
Survey (GOODS) ACS (\cite{giavalisco}) observed optical morphologies of
optically-bright power-law galaxies in the CDF-S. 
 The power-law galaxies spectroscopically classified 
as BLAGN (and SED type) display bright nuclear point sources suggesting that
 the optical light is dominated by the AGN, which is consistent with  the fact
 that our power-law criteria selects the most X-ray luminous AGN  (see
also \cite{rigby06}). The power-law galaxies without broad lines (those
classified as HEX or LEX by \cite{szokoly})
have disky or irregular morphologies. A morphological characterization of the IR power-law
galaxies not detected in X-rays is difficult as only about half of them 
are detected in the GOODS/ACS images and they are faint. 
As can be seen for a few examples in
Fig.~4 they have irregular, 
knotty, and/or interacting morphologies, and 
do not appear to contain bright point sources.

\section{Obscuration and Obscured Fraction}
In the distant universe the X-ray background and luminosity synthesis models
predict global obscured ($N_{\rm H} \ge 10^{22}\,{\rm  cm}^{-2}$) to
unobscured ratios of 3:1 to 4:1 (e.g., \cite{comastri,gilli}), significantly 
higher than the
observed ratios of spectroscopically identified X-ray sources in deep fields
(e.g., \cite{barger,szokoly}) including those detected in the mid-IR (e.g.,
\cite{rigby04}). \cite{donley06} estimated the intrinsic 
column densities of each of the X-ray well detected and weakly X-ray emitting
power-law galaxies in the CDF-N. We also included a few CDF-S IR power-law galaxies 
(\cite{aah06}) for which \cite{rigby06} estimated the X-ray column
densities. The column density
distribution of the power-law galaxies (Fig.~7) 
is significantly different from that of optically bright X-ray AGN with
spectroscopic classifications (see also \cite{tozzi}), including
the optically-dull AGN which are believed to 
suffer strong obscuration (\cite{rigby06}). 
From Fig.~7 it is clear that the weakly
detected IR power-law galaxies are consistent with being obscured ($N_{\rm H} \sim
10^{22}-10^{24}\,{\rm  cm}^{-2}$) but not Compton-thick ($N_{\rm H} \ge
10^{24}\,{\rm  cm}^{-2}$). If all the X-ray non-detected power-law
galaxies are obscured, the maximum obscured ratio is 4:1 (for the CDF-N
power-law galaxies). This $N_{\rm H}$-based obscured
fraction of power-law galaxies agrees well with the ratio of BLAGN (unobscured)
SED vs. NLAGN (obscured) SEDs found in the CDF-S.

We can finally estimate the ratio of obscured to unobscured mid-IR 
detected AGN in the CDF-S. The unobscured AGN
are all those X-ray sources (detected in the hard band 
to make sure they are AGN) with $N_{\rm H}<10^{22}\,{\rm cm}^{-2}$, 
whereas in the obscured category we include all obscured X-ray 
sources with $N_{\rm H}>10^{22}\,{\rm cm}^{-2}$, and all the obscured 
IR power-law galaxies. We find an
observed ratio of obscured to unobscured AGN of 2:1 in the CDF-S. 
Comparing with the predictions of \cite{treister} for the $24\,\mu$m detected
AGN number density we find that our sample of power-law galaxies only accounts for
approximately 20\% of all the mid-IR emitting obscured AGN in the CDF-N. This
fraction can be as high as $\sim 50\%$ in the CDF-S as a result of
larger sample of power-law galaxies there (we did not impose IRAC high
S/N detections, see \S2 and \cite{aah06,donley06} for
details). The remainder should 
have SEDs dominated by or strongly
affected by the host galaxy or red power-law SEDs that fall below the IRAC
detection limit. This is not surprising as our power-law criteria require the
AGN to be energetically dominant in the near to mid-IR.

\section{Other Infrared-Based Searches for Obscured AGN}

The selection of IR-bright optically-faint galaxies has been suggested as another
method for identifying obscured AGN (\cite{houck05,weedman06}), although the
selection criteria ($R>23.9\,$ and $f_{24\,\mu{\rm m}} > 0.75-1\,$mJy) were
  set so that IRS follow-up spectroscopy could be obtained. These criteria select
  mostly high-$z$ (median $z \sim 2.2$) galaxies, with only a small fraction
  showing the characteristic aromatic feature  emission of star formating galaxies. The majority 
have IRS spectra similar to local AGN-dominated ULIRGs with either a featureless
power-law rest-frame mid-IR continuum or deep silicate features at
$9.7\,\mu$m. In addition, their 
SEDs lack a strong $1.6\,\mu$m stellar bump, and are similar to those of IR power-law
galaxies. Their properties are consistent with being optically obscured AGN-powered ULIRGs
with $L_{\rm IR} >10^{12}\,{\rm L}_\odot$ (see \cite{houck05,weedman06} for details). 

Only a  few galaxies in the CDF-N and
CDF-S fall within the \cite{houck05,weedman06} flux density cuts. 
We can instead compare with the 
$24\,\mu$m/$8\,\mu$m vs. $24\,\mu$m/$R$-band diagram criterion proposed by \cite{yan04,yan}  
to select obscured AGN. 
The location of our CDF-N power-law galaxies on this diagram can be seen in
Fig.~8. We also show the comparison X-ray 
sources (from \cite{alexander}) and other IRAC sources in the field that do
not meet the power-law criteria. We find that all these 
samples cover a large range in colors, but
the power-law galaxies comprise a significant fraction ($\sim 30-40\%$) of the
highly optically reddened members of the comparison X-ray sample. This
suggests that the power-law selection is capable of detecting both optically
obscured and unobscured AGN, and that a large fraction of the
IR-bright/optically-faint sources in the comparison sample have power-law SEDs
in the near and mid-IR (Fig.~8).

Radio emission is another good way to select AGN as it is unaffected by dust
absorption. \cite{donley05} in the CDF-N used a
radio to mid-IR ratio to select galaxies that are too bright in radio to be
star-forming galaxies. They found that $\sim 30$\% of their radio-loud AGN
are not detected in X-rays suggesting strong
obscuration. \cite{martinez05} looked for a population of radio intermediate
and radio quiet AGN by 
selecting $24\,\mu$m sources
($f_{24\,\mu{\rm m}} \sim 0.3-1\,$mJy)  with radio
emission and imposed a flux density cut at $3.6\,\mu$m to filter out type-1
and radio-loud QSOs. They found a population of 
QSOs at $1.4<z<4.2$ (median $z=2$,
the epoch of QSO maximum activity) with a ratio of obscured  to
unobscured of ($2-3$):1, and postulated that this population of obscured AGN may
be responsible for most of the black hole growth in the young universe.


\begin{figure}[!t]
\vspace{-1cm}
\includegraphics[width=12cm]{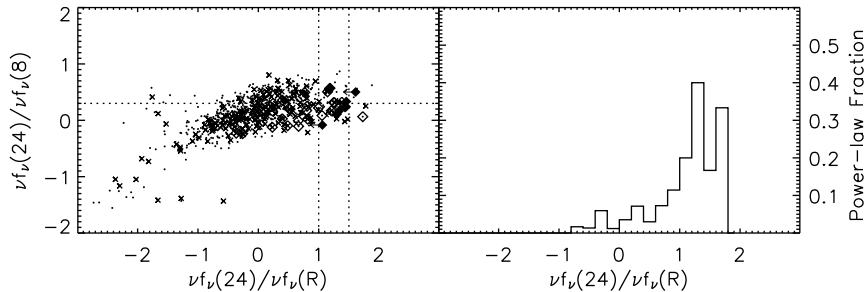}
\vspace{-7.5cm}
\caption{Location of the CDF-N power-law galaxies and comparison samples (symbols as
in Fig.~1) 
 on the color-color diagram of \cite{yan04} (left panel) where the
dotted lines indicate the Yan et al. selection criteria for possible
dust-reddened AGN. The right panel shows the
power-law fraction of the IRAC comparison sample (non power-law galaxies) 
as a function of the $24\,\mu$m
over $R$-band flux ratio.}
\label{fig:8}       
\end{figure}

\section{Summary}

Deep X-ray cosmological surveys are efficient at detecting AGN at high-$z$
but they only account for a $\sim$ 1:1 ratio  of obscured-to-unobscured AGN,
whereas synthesis models of the X-ray background require ratios of between  
3:1 and 4:1. 
Thus there is a significant population of obscured ($N_{\rm H} >
10^{23}-10^{24}\,{\rm cm}^{-2}$) AGN being missed by current
deep X-ray ($<10\,$keV) observations. We describe searches for this population.
Our selection criterion is based on the characteristic IR 
power-law emission shown by local QSOs. By selecting galaxies with 
power-law emission in the {\it Spitzer}/IRAC
bands ($3.6, \, 4.5, \, 5.8, \, {\rm and \, } 8\,\mu$m) we 
avoid high-$z$ galaxies whose SEDs are dominated by
stellar emission and star formation peaking at $1.6\,\mu$m. 

Only $\sim 50$\% of IR power-law galaxies are detected in deep ($1-2\,$Ms)
X-ray exposures. This fraction increases to 75\% if we include weakly detected
X-ray sources at the $3\,\sigma$ level in the field with the deepest X-ray
exposure (CDF-N). The optical (faint), IR (mostly ULIRGs and hyper luminous IR galaxies), 
and X-ray properties, the X-ray 
column densities $N_{\rm H}$  
(moderately obscured, but not Compton-thick) and redshift ($z>1$)
distributions, and SED shapes of a large fraction (up to $80\%$) of IR
power-law galaxies are
significantly different from bright X-ray selected AGN. This may indicate  
that a large fraction of IR power-law galaxies are good candidates to
host obscured AGN,
and could account for a ratio of 2:1 of obscured-to-unobscured AGN at high-$z$.  

Other mid-IR based criteria
(e.g., \cite{donley05,houck05,lacy04,lacy06,martinez05,stern05,yan04,yan,weedman06})
are also finding
populations of obscured AGN. There might be a significant overlap
between populations of bright mid-IR  AGN selected with all these methods, and thus 
a complete census of the {\it  entire} obscured AGN population is needed to
determine whether we can account for the obscured fraction of the
 X-ray background.

$\,$

A. A. H. acknowledges support from the Spanish 
Plan Nacional del Espacio under grant ESP2005-01480
and P.~G. P.-G. from the Spanish Programa Nacional de Astronom\'{\i}a y
Astrof\'{\i}sica under grant AYA 2004-01676 and the Comunidad de Madrid
ASTRID I+D project. Support for this work was also provided by
NASA through Contract no. 
960785 and 1256790 issued by JPL/Caltech.



\printindex

\end{document}